\documentclass[smallextended]{svjour3}       
\smartqed  
\usepackage{graphicx,epsfig}

\begin{document}

\title{
Bimodal behavior of post-measured entropy and one-way quantum deficit for two-qubit X states
}

\author{
M.A.Yurischev
}

\institute{
M.~A.~Yurischev
\at
Institute of Problems of Chemical Physics, Russian Academy of Sciences,
Chernogolovka, Moscow Region, Russia 142432\\     
\email{yur@itp.ac.ru} }

\date{Received:}

\maketitle

\begin{abstract}
A method for calculating the one-way quantum deficit is developed.
It involves a careful study of post-measured entropy shapes.
We discovered that in some regions of X-state space the post-measured entropy $\tilde S$
as a function of measurement angle $\theta\in[0,\pi/2]$ exhibits a bimodal behavior inside
the open interval $(0,\pi/2)$, i.e., it has two interior extrema: one minimum and one maximum.
Furthermore, cases are found when the interior minimum of such a bimodal function
$\tilde S(\theta)$ is less than that one at the endpoint $\theta=0$ or $\pi/2$.
This leads to the formation of a boundary between the phases of one-way quantum deficit via
{\em finite} jumps of optimal measured angle from the endpoint to the interior minimum.
Phase diagram is built up for a two-parameter family of X states.
The subregions with variable optimal measured angle
are around 1$\%$ of the total region, with their relative linear sizes achieving $17.5\%$,
and the fidelity between the states of those subregions can be reduced to $F=0.968$.
In addition, a correction to the one-way deficit due to the interior minimum can achieve
$2.3\%$.
Such conditions are favorable to detect the subregions with variable optimal measured angle
of one-way quantum deficit in an experiment.

%
%
\end{abstract}

\section{Introduction}
\label{sect:Intro}
Quantum correlation is a key feature of quantum mechanics and it lies at the heart
of quantum information science.
Besides the quantum entanglement and discord, the one-way quantum deficit is one
of the most important measures of quantum correlation
\cite{MBCPV12,Str15,ABC16,BDSRSS17}.
The entanglement is identical to the discord and one-way deficit for
the pure quantum states,
whereas the discord and one-way deficit coincide in considerably more
general cases --- they are the same for the Bell-diagonal states and
even for the X states with zero Bloch vector for one qubit (i.e., with a single maximally
mixed marginal) if the local measurements are performed on this qubit \cite{YF16}.

Definitions of quantum discord $Q$ and one-way quantum deficit $\rm\Delta$ involve
the minimization procedure to obtain the optimal measurement performed on one part
of bipartite system.
This procedure for the two-qubit systems with X density matrix is reduced to the
minimization problem on one variable -- the polar angle $\theta\in[0,\pi/2]$
(see Refs.~\cite{CRC10,VR12,JY16,WJFWCF15}).
Moreover, a formula for the quantum discord is presented in a partially analytic
(piecewise-analytical-numerical) form \cite{Y14,Y14a,Y15},
\begin{equation}
   \label{eq:Q3}
   Q=\min\{Q_0, Q_{\vartheta}, Q_{\pi/2}\}.
\end{equation}
Here, the subfunctions (branches) $Q_0$ and $Q_{\pi/2}$ are the analytical expressions
(corresponding to the discord with optimal measurement angles equaling zero and $\pi/2$,
respectively) and only the third branch $Q_{\vartheta}$ requires to perform numerical
minimization to obtain state-dependent minimizing angle $\vartheta\in(0,\pi/2)$ if, of course,
the interior minimum exists.
Equations for 0- and $\pi/2$-boundaries separating respectively the $Q_0$ and $Q_{\pi/2}$
regions with the $Q_{\vartheta}$ one can be written as \cite{Y14,Y14a,Y15}
\begin{equation}
   \label{eq:QII}
   Q^{\prime\prime}(0)=0, \qquad   Q^{\prime\prime}(\pi/2)=0.
\end{equation}
Here $Q^{\prime\prime}(0)$ and $Q^{\prime\prime}(\pi/2)$ are the second derivatives
of the measurement-dependent discord function $Q(\theta)$ with respect to  $\theta$
at the endpoints $\theta=0$ and $\pi/2$, correspondingly.
The equations (\ref{eq:QII}) are based on the unimodality hypothesis for the
function $Q(\theta)$ which is confirmed for different classes of X states
\cite{Y15,Y17}.
Notice that Eqs.~(\ref{eq:QII}) reflect the bifurcation mechanism of appearance of the minimum
inside the interval $(0,\pi/2)$.

On the other hand, as mentioned above, there is a close connection between the one-way
quantum deficit and quantum discord.
Therefore it would be tempting to propose that similar properties are valid for the
measurement-dependent one-way quantum deficit function $\Delta(\theta)={\tilde S}(\theta)-S$,
where $S$ is the pre-measurement entropy.

Recently, the authors \cite{YWF16} have claimed the result which is reduced to the statement
that the one-way quantum deficit ${\rm\Delta}=\min_\theta\Delta(\theta)$ for the general
X states is given by
\begin{equation}
   \label{eq:D2}
   \rm\Delta=\cases{\Delta(\vartheta),\qquad\qquad\qquad\ \ \Delta^{\prime\prime}(0)<0\ {\rm and}\
	 \Delta^{\prime\prime}(\pi/2)<0,\ \vartheta\in(0,\pi/2);\cr
	 \min\{\Delta(0), \Delta(\pi/2)\},\  {\rm others}.}
\end{equation}
If the function $\Delta(\theta)$ is monotonic or has single extremum inside the interval $(0,\pi/2)$
this conclusion takes place.

In the present paper we show that the post-measured entropy and consequently
the measurement-dependent one-way quantum deficit can display more general behavior which
refutes the relation (\ref{eq:D2}).
We discuss the difficulties arisen from a new type of behavior and propose,
instead of Eq.~(\ref{eq:D2}), the method giving the correct calculation of one-way
deficit for two-qubit X states.

\section{Results and discussion}
\label{sect:ResDis}
Let us consider a two-parameter family of X states
\begin{equation}
   \label{eq:rho}
   \rho_{AB}=q_1|\Psi^+\rangle\langle\Psi^+| + q_2|\Psi^-\rangle\langle\Psi^-|
	 + (1-q_1-q_2)|00\rangle\langle 00|,
\end{equation}
where $|\Psi^{\pm}\rangle=(|01\rangle\pm|10\rangle)/\sqrt{2}$.
This family generalizes the class of special X states from Ref.~\cite{YWF16}
which corresponds to the case $q_1=0$.

\begin{figure}[t]
\begin{center}
\epsfig{file=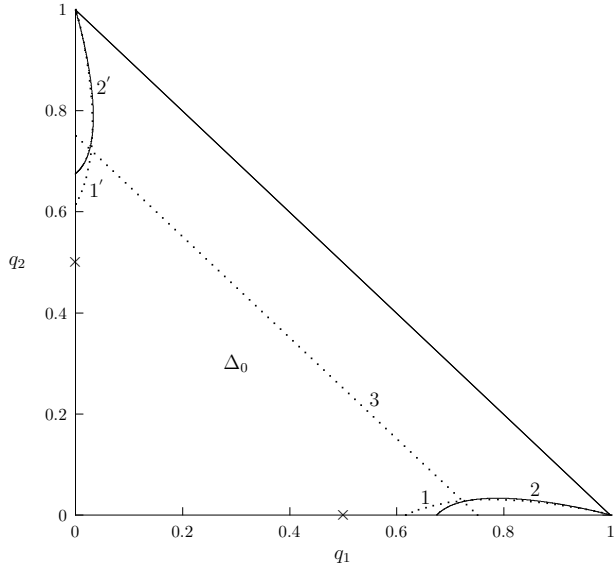,width=8cm}
\caption{
Triangle ${\cal T}$ in the plane $(q_1,q_2)$ with vertices $(0,0)$, $(0,1)$, and $(1,0)$
is the permitted region for the parameters $q_1$ and $q_2$.
Dotted lines 1 and 1$^{'}$ are the boundaries defined by the equation $\Delta_0=\Delta_{\pi/2}$.
Solid lines 2 and 2$^{'}$ are the $\pi/2$-boundaries.
Dotted line 3 is the path $q_1+q_2=0.75$.
Crosses ($\times$) at the points $(0,0.5)$ and $(0.5,0)$ mark the 0-boundaries
}
\label{fig:zpd4}
\end{center}
\end{figure}

The density matrix (\ref{eq:rho}) in open form is given as
\begin{equation}
   \label{eq:rho1}
   \rho_{AB}
	 =\left(
      \begin{array}{cccc}
      1-q_1-q_2&0&0&0\\
      0&(q_1+q_2)/2\ &(q_1-q_2)/2\ &0\\
      0&(q_1-q_2)/2\ &(q_1+q_2)/2\ &0\\
      0&0&0&0
      \end{array}
   \right).
\end{equation}
Eigenvalues of this matrix equal
\begin{equation}
   \label{eq:lam}
   \lambda_1=1-q_1-q_2,\quad \lambda_2=q_1,\quad \lambda_3=q_2,\quad \lambda_4=0.
\end{equation}
Owing to the non-negativity requirement for any density matrix, one obtains that the domain
of definition for the parameters (arguments) $q_1$ and $q_2$ is restricted by conditions
\begin{equation}
   \label{eq:q1q2}
   q_1\ge0,\quad q_2\ge0,\quad q_1+q_2\le1.
\end{equation}
Thus, the domain in plane $(q_1,q_2)$ is the triangle ${\cal T}$ which is shown in
Fig.~\ref{fig:zpd4}.

One-way quantum deficit (quantum work deficit) for a bipartite state $\rho_{AB}$ is defined
as the minimal increase
of entropy after a von Neumann measurement on one party (without loss of generality, say, $B$)
\cite{SKB11,CPBSS15,CDSPSS15}
\begin{equation}
   \label{eq:tildeD}
   {\rm\Delta}=\min_{\{\rm\Pi_k\}}S(\tilde\rho_{AB})-S(\rho_{AB}),
\end{equation}
where
\begin{equation}
   \label{eq:rho_tilde}
   \tilde\rho_{AB}=\sum_k(I\otimes{\rm\Pi}_k)\rho_{AB}(I\otimes{\rm\Pi}_k)^+
\end{equation}
is the weighted average of post-measured states and $S(\cdot)$ means the von
Neumann entropy.
In Eqs.~(\ref{eq:tildeD}) and (\ref{eq:rho_tilde}), $\rm\Pi_k$ ($k=0,1$) are the general
orthogonal projectors
\begin{equation}
   \label{eq:Pi}
   {\rm\Pi}_k=V\pi_kV^+,
\end{equation}
where $\pi_k=|k\rangle\langle k|$ and transformations $\{V\}$ belong to
the special unitary group $SU_2$.
Rotations $V$ may by parametrized by two angles $\theta$ and $\phi$
(polar and azimuthal, respectively):
\begin{equation}
   \label{eq:V}
   V
	 =\left(
      \begin{array}{cc}
      \cos(\theta/2)&-e^{-i\phi}\sin(\theta/2)\\
      e^{i\phi}\sin(\theta/2)&\cos(\theta/2)
      \end{array}
   \right)
\end{equation}
with $0\le\theta\le\pi$ and $0\le\phi<2\pi$.

Using Eq.~(\ref{eq:lam}) one gets the pre-measured entropy
\begin{equation}
   \label{eq:preS}
   S(q_1,q_2)\equiv S(\rho_{AB})=-q_1\log{q_1}-q_2\log{q_2}-(1-q_1-q_2)\log{(1-q_1-q_2)}.
\end{equation}

Eigenvalues of the matrix ${\tilde\rho}_{AB}$ are equal to
\begin{eqnarray}
   \label{eq:Lam}
	 \Lambda_{1,2}&&=\frac{1}{4}\lbrack\!\lbrack1+(1-q_1-q_2)\cos\theta\pm\{[1-q_1-q_2+(1-2q_1-2q_2)\cos\theta]^2
	 \nonumber\\
	 &&+(q_1-q_2)^2\sin^2\theta\}^{1/2}\rbrack\!\rbrack
	 \nonumber\\
	 \\
	 \Lambda_{3,4}&&=\frac{1}{4}\lbrack\!\lbrack1-(1-q_1-q_2)\cos\theta\pm\{[1-q_1-q_2-(1-2q_1-2q_2)\cos\theta]^2
	 \nonumber\\
	 &&+(q_1-q_2)^2\sin^2\theta\}^{1/2}\rbrack\!\rbrack.
	 \nonumber
\end{eqnarray}
It is seen that the azimuthal angle $\phi$ has dropped out from the given expressions.
This is due to the fact that one pair of non-diagonal entries of the density matrix (\ref{eq:rho1})
vanishes.
Using Eqs.~(\ref{eq:Lam}) we arrive at the post-measured entropy (entropy after measurement)
\begin{equation}
   \label{eq:postS}
   \tilde S(\theta;q_1,q_2)\equiv S(\tilde\rho_{AB})=h_4(\Lambda_1,\Lambda_2,\Lambda_3,\Lambda_4),
\end{equation}
where $h_4(x_1,x_2,x_3,x_4)=-\sum_{i=1}^4x_i\log x_i$ with additional condition
$x_1+x_2+x_3+x_4=1$ is the quaternary entropy function.

Notice that function $\tilde S$ of argument $\theta$ is invariant under
the transformation $\theta\to\pi-\theta$ therefore it is enough to restrict oneself by
values of $\theta\in[0,\pi/2]$.
Moreover, the pre- and post-measured entropies $S$ and $\tilde S$, as functions
of $q_1$ and $q_2$, are symmetric under the exchange
$q_1\rightleftharpoons q_2$.

Equations~(\ref{eq:preS})--(\ref{eq:postS}) define the measurement-dependent one-way
deficit function $\Delta(\theta)={\tilde S}(\theta)-S$.
Direct calculations show that for every choice of model parameters the function
${\tilde S}(\theta)$ and hence $\Delta(\theta)$ possess an important property,
namely their first derivatives with respect to $\theta$ identically equal zero at
both endpoints $\theta=0$ and $\theta=\pi/2$:
\begin{equation}
   \label{eq:postSD1}
   {\tilde S}^{\prime}(0)=\Delta^{\prime}(0)\equiv0,\qquad
   {\tilde S}^{\prime}(\pi/2)=\Delta^{\prime}(\pi/2)\equiv0.
\end{equation}

From Eqs.~(\ref{eq:Lam}) and (\ref{eq:postS}) we get the expressions for the post-measurement entropy
at the endpoint $\theta=0$,
\begin{equation}
   \label{eq:S0}
   \tilde S_0(q_1,q_2)=-(1-q_1-q_2)\log(1-q_1-q_2)-(q_1+q_2)\log[(q_1+q_2)/2],
\end{equation}
and at the second endpoint $\theta=\pi/2$:
\begin{equation}
   \label{eq:S1}
   \tilde S_{\pi/2}(q_1,q_2)=\log2+h((1+\sqrt{(1-q_1-q_2)^2+(q_1-q_2)^2})/2),
\end{equation}
where $h(x)=-x\log x-(1-x)\log(1-x)$ is the Shannon binary entropy function.
Together with Eq.~(\ref{eq:preS}) they supply us with explicit expressions for the
one-way deficit at the endpoints:
$\Delta_0=\Delta(0)$ and $\Delta_{\pi/2}=\Delta(\pi/2)$.
In particular, 
if $q_1$ or $q_2$ equals zero then $\Delta_0=q\log2~(=q, {\rm bit})$,
where $q=\{q_1,q_2\}$.

Solving the transcendental equation
\begin{equation}
   \label{eq:D0D1}
   \Delta_0=\Delta_{\pi/2}
\end{equation}
or, the same,
$\tilde S_0=\tilde S_{\pi/2}$ we find the subregions in the plane $(q_1,q_2)$,
where $\Delta_{\pi/2}<\Delta_0$ (restricted in Fig.~\ref{fig:zpd4} by dotted
curves 1 and 1$^{\prime}$ and corresponding Cartesian axes $Oq_1$ and $Oq_2$)
and where, v.v., $\Delta_0<\Delta_{\pi/2}$ (marked in Fig.~\ref{fig:zpd4} by symbol
$\Delta_0$).
The curve 1 has two endpoints on the axis $Oq_1$: at $q_1=0.61554$ and $q_1=1$.
Analogously for the curve 1$^{\prime}$ (see Fig.~\ref{fig:zpd4}).

The 0- and $\pi/2$-boundaries, i.e., where respectively the second derivatives
\begin{equation}
   \label{eq:D11}
   \Delta^{\prime\prime}(0)=0\quad
   {\rm and}\quad
   \Delta^{\prime\prime}(\pi/2)=0
\end{equation}
or, the same,
$\tilde S^{\prime\prime}(0)=0$ and $\tilde S^{\prime\prime}(\pi/2)=0$,
will be needed below.
As calculations yield,
\begin{eqnarray}
   \label{eq:S11P}
	 \tilde S^{\prime\prime}(\pi/2)&&=\frac{(q_1-q_2)^2}{2r^3}[r^2-(1-2q_1-2q_2)^2]\ln\frac{1+r}{1-r}
	 \nonumber\\
	 &&-\frac{(1-q_1-q_2)^2}{1-r^2}[1-2(1-2q_1-2q_2)(1-\frac{1-2q_1-2q_2}{2r^2})],
\end{eqnarray}
where
\begin{equation}
   \label{eq:r}
   r=\sqrt{(1-q_1-q_2)^2+(q_1-q_2)^2}.
\end{equation}
On the other hand, calculations show that the second derivative
$ \tilde S^{\prime\prime}(\theta)$ with respect to $\theta$
is finite at $\theta=0$ only when $q_1q_2=0$:
\begin{equation}
   \label{eq:S110}
   \tilde S^{\prime\prime}(0)=\frac{1-3q+2q^2}{2-3q}\ln\frac{2(1-q)}{q},
\end{equation}
where again $q=\{q_1,q_2\}$.
The roots of equation $\tilde S^{\prime\prime}(0)=0$ are 1/2 and 1.
Thus, the bifurcation 0-boundary exists only if $q_1=0$
or, inversely, $q_2=0$ (that is, only at two points on each of the Cartesian axes
$Oq_1$ and $Oq_2$).
The corresponding 0-boundaries $q_1=1/2$, when $q_2=0$, and $q_2=1/2$, when $q_1=0$
are shown in Fig.~\ref{fig:zpd4} by the crosses.

The results of numerical solution of the equation $\tilde S^{\prime\prime}(\pi/2)=0$
are presented in Fig.~\ref{fig:zpd4} by solid lines 2 and 2$^{\prime}$.
The endpoints for the curve 2 on the axis $Oq_1$ are $q_1=0.67515$ and $q_1=1$.
The curves 1 and 2 intersect at the point with coordinates $q_1=0.739\,409$ and
$q_2=0.029\,686$ ($q_1+q_2=0.769\,095$).
Analogously for the curves 1$^{\prime}$ and 2$^{\prime}$ with, of course,
permutation of $q_1$ and $q_2$ (see again Fig.~\ref{fig:zpd4}).

\begin{figure}[t]
\begin{center}
\epsfig{file=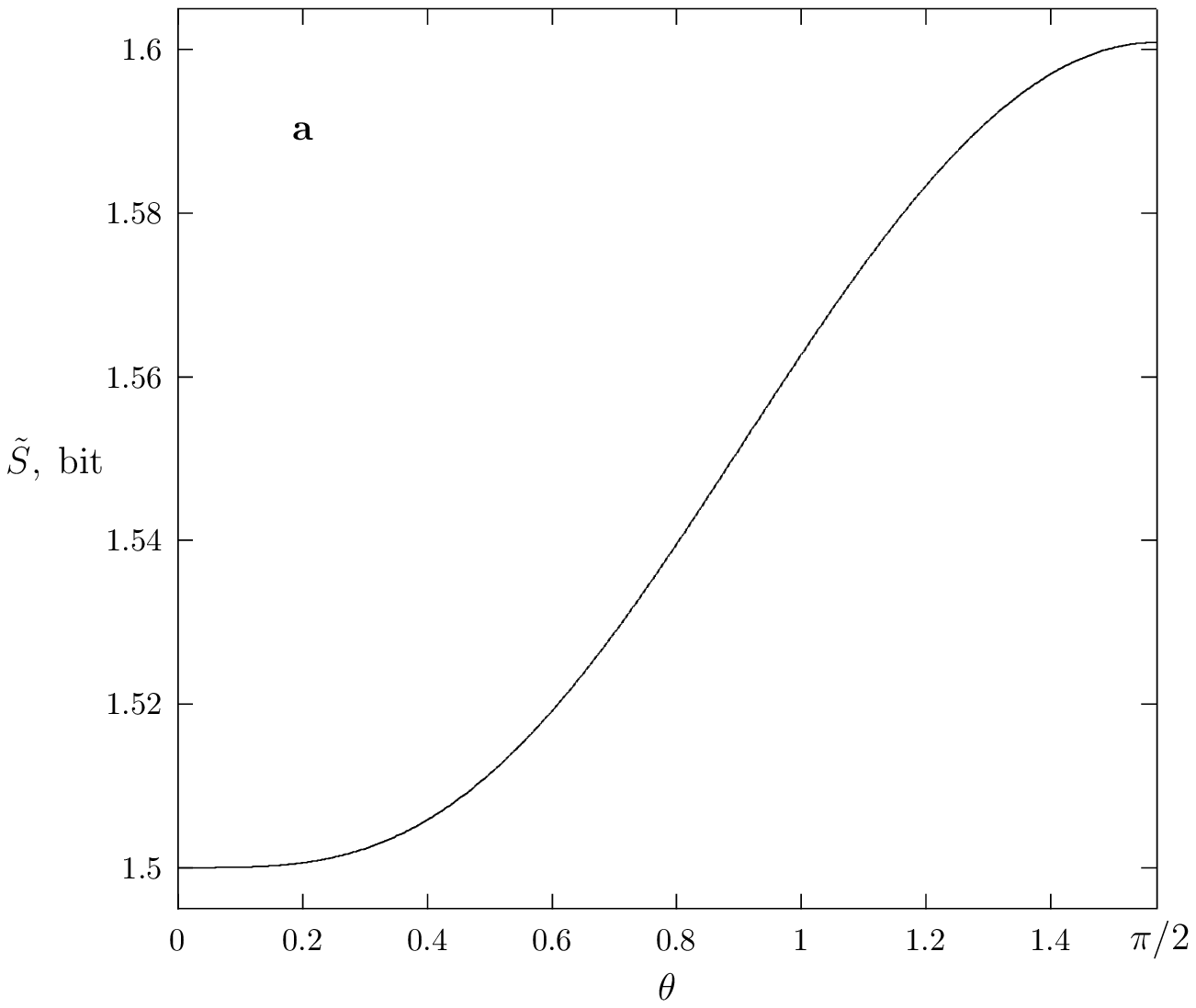,width=5.2cm}
\vspace{1cm}
\hspace{1cm}
\epsfig{file=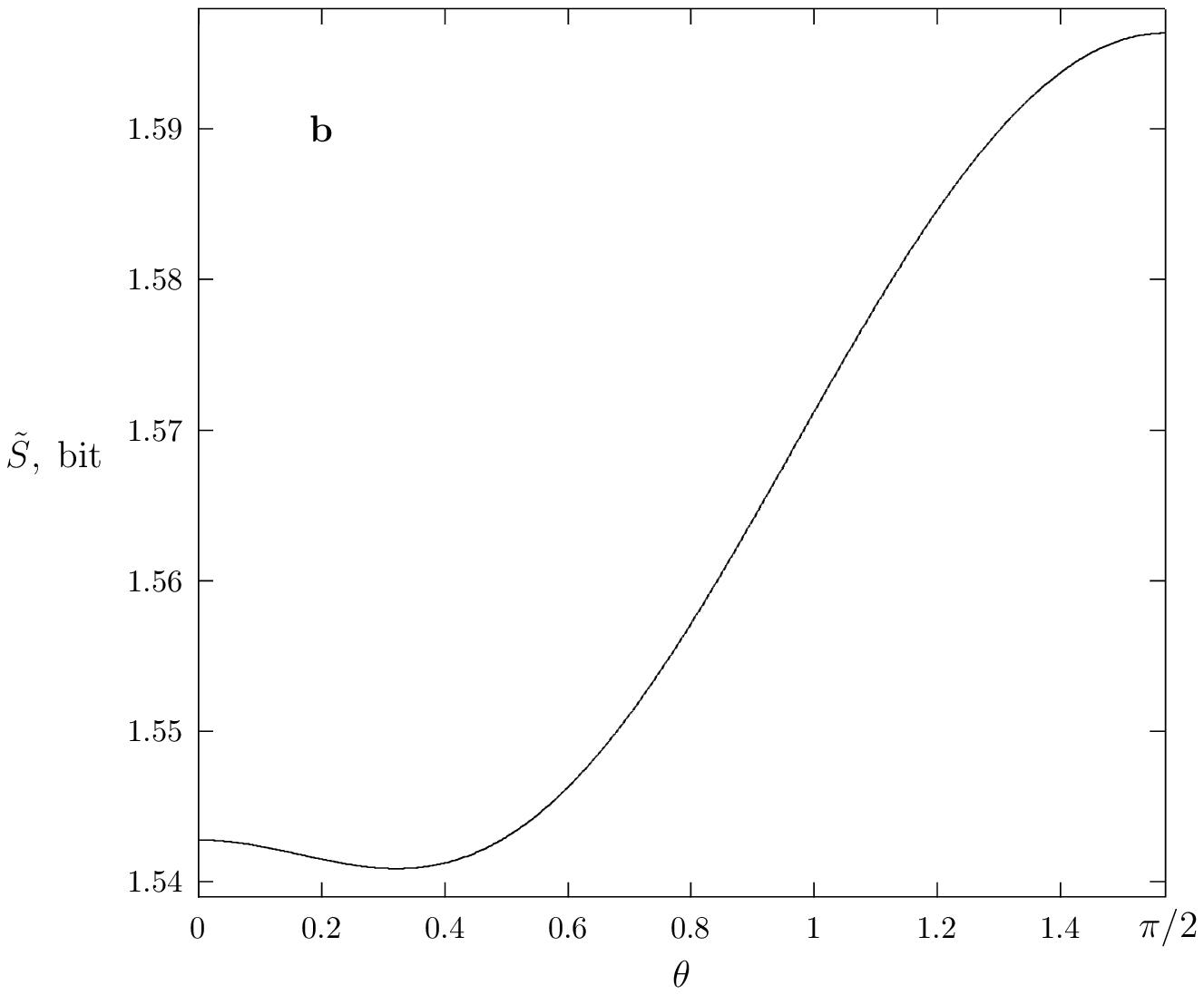,width=5.2cm}
\epsfig{file=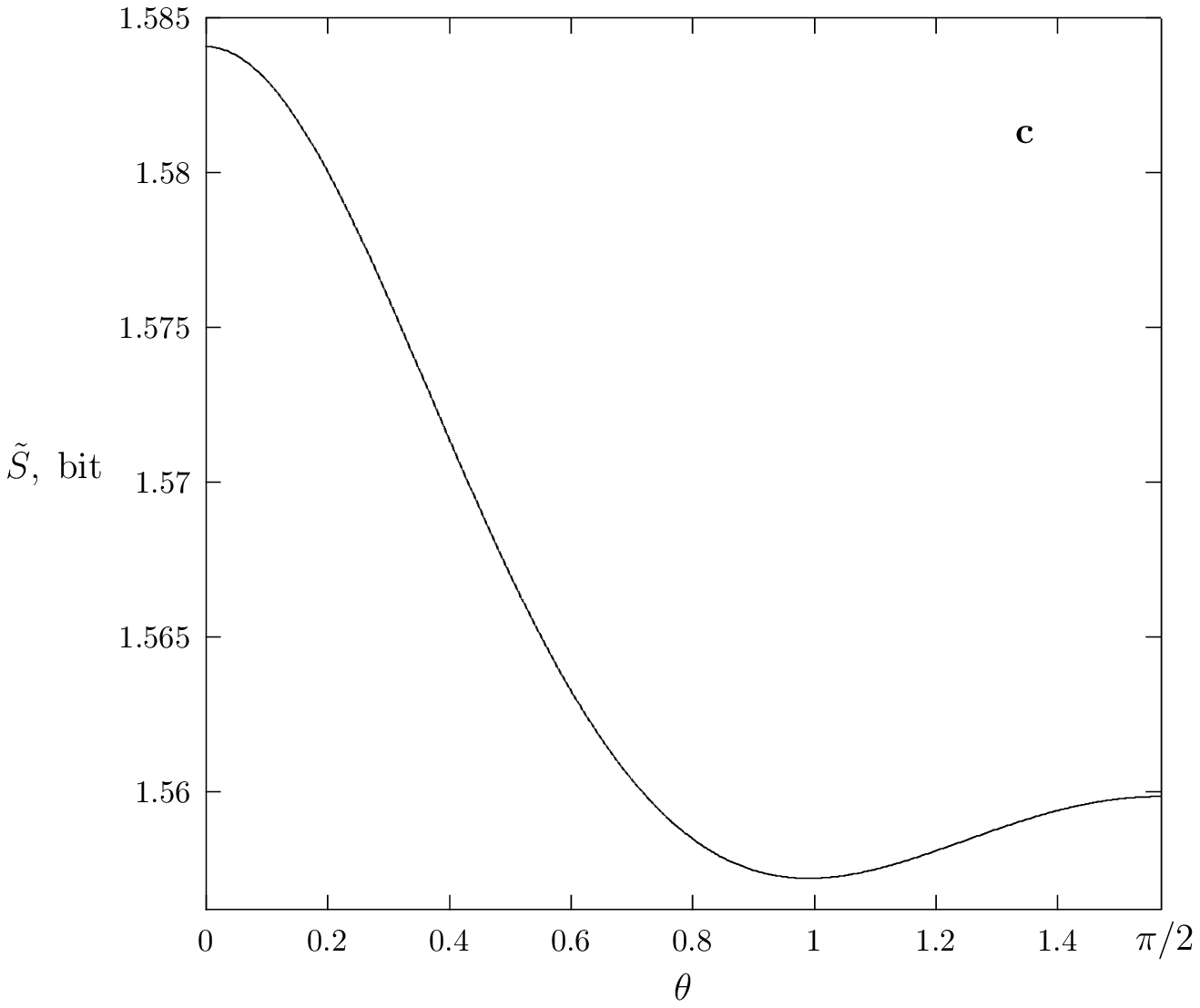,width=5.2cm}
\vspace{1cm}
\hspace{1cm}
\epsfig{file=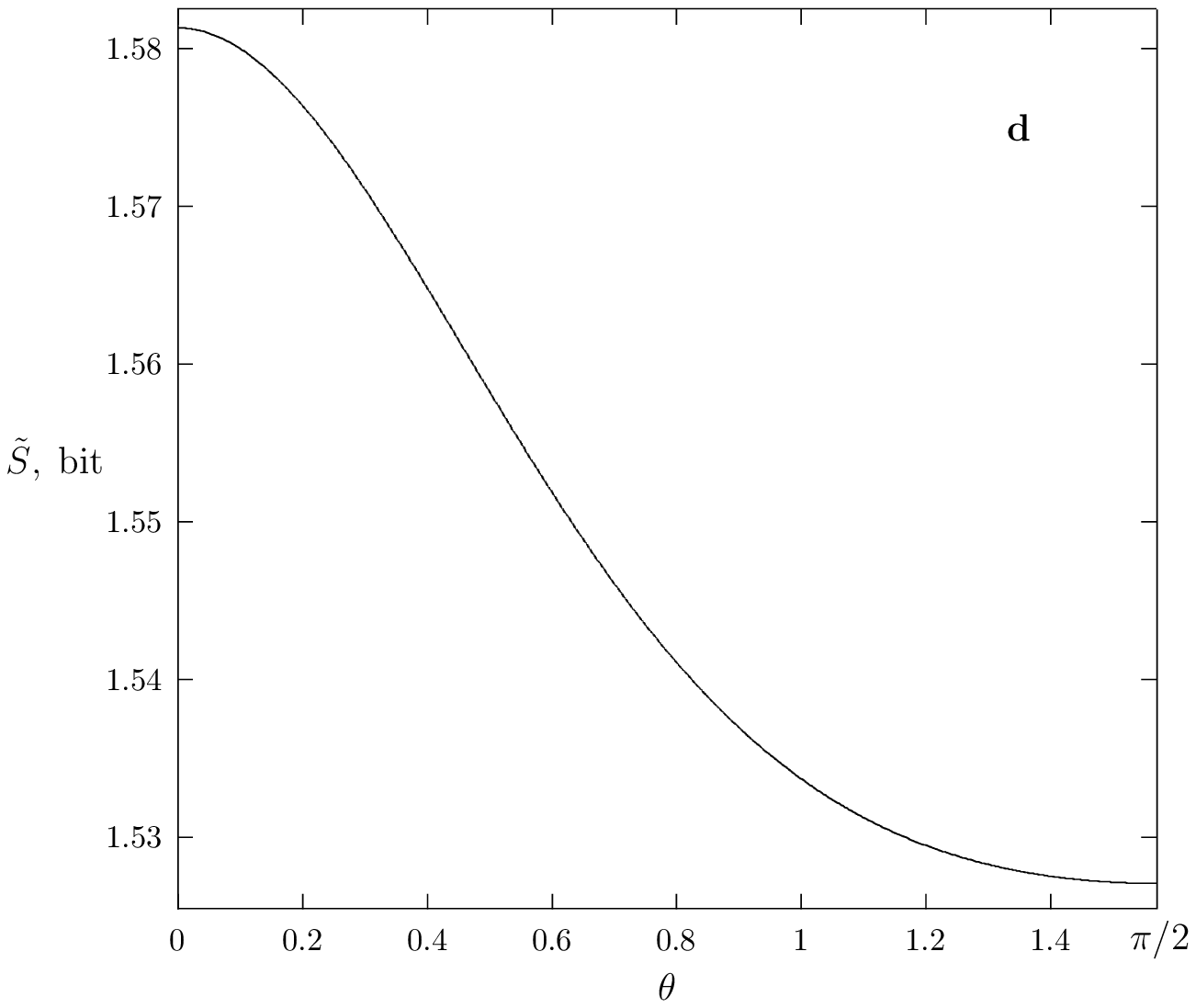,width=5.2cm}
\caption{
Post-measurement entropy $\tilde S$ vs $\theta$ by $q_2=0$ and $q_1=0.5$~(a),
0.55~(b), 0.65~(c), and 0.7~(d)
}
\label{fig:zs4}
\end{center}
\end{figure}
Let us consider the behavior of post-measured entropy $\tilde S(\theta)$
and non-minimized one-way deficit $\Delta(\theta)$ by moving along different
trajectories (paths) in the triangle ${\cal T}$.

Start with the passing along the leg of triangle ${\cal T}$.
Figure~\ref{fig:zs4} shows the evolution of shape of the post-measured entropy
$\tilde S(\theta; q_1, 0)$ with changing the parameter $q_1$.
The curve has the monotonically increasing behavior when the argument $q_1$ varies from
$q_1=0$ to $q_1=1/2$; see Fig.~\ref{fig:zs4}(a).
At the point $q_1=1/2$ a bifurcation of the minimum at $\theta=0$ occurs.
Then, when $q_1$ increases from 0.5 to 0.67515, the curve $\tilde S(\theta)$
has, as shown in Figs.~\ref{fig:zs4}(b) and (c), the interior minimum,
with the function $\tilde S(\theta)$ being here unimodal.
So, the region with variable optimal angle $\vartheta$ takes up a part $0.17515\approx17.5\%$
on the section $[0,1]$ of $Oq_1$ axis and the fidelity of states at points $(0.5,0)$ and
$(0.67515,0)$ is equal to $F=96.8\%$\footnote{
   Note for comparison that in two-photon experiments one achieves now the values of fidelity
   $F=99.8(2)\%$ \cite{BSPBG13} and $F=99.8(1)\%$ \cite{Guo16}.
}.
The position of such a local minimum smoothly increases from zero to $\pi/2$;
see again the curves in Figs.~\ref{fig:zs4}(b) and (c).
The values of $\tilde S_0$ and $\tilde S_{\pi/2}$ become equal
at the point $q_1=0.61554$ ($\tilde S_0=\tilde S_{\pi/2}=1.57667$~bit,
hence $\Delta_{\pi/2}=\Delta_0=q_1=0.61554$~bit)
and the depth of interior minimum is 0.01397~bit what gives a relative correction
to the one-way deficit equaled $\delta{\rm\Delta}=2.3\%$.
Then, at the value of $q_1=0.67515$, the system experiences a new sudden transition -- from
the branch, which is characterized by the continuously changing optimal angle $\vartheta$
in the full interval (from 0 to $\pi/2$), to the branch $\tilde S_{\pi/2}$ with constant
optimal measurement angle equaled $\pi/2$.
After this the curves of post-measured entropy exhibit monotonically decreasing behavior
as illustrated in Fig.~\ref{fig:zs4}(d).
One should emphasize here that the minimized one-way quantum deficit,
${\rm\Delta}=\min_\theta\Delta(\theta)$, vs the model parameter $q_1$ is continuous and smooth.
Nevertheless, the function ${\rm\Delta}(q_1)$ has nonanalyticities at the points $q=0.5$ and 0.67515
which manifest themselves in higher derivatives.

\begin{figure}[t]
\begin{center}
\epsfig{file=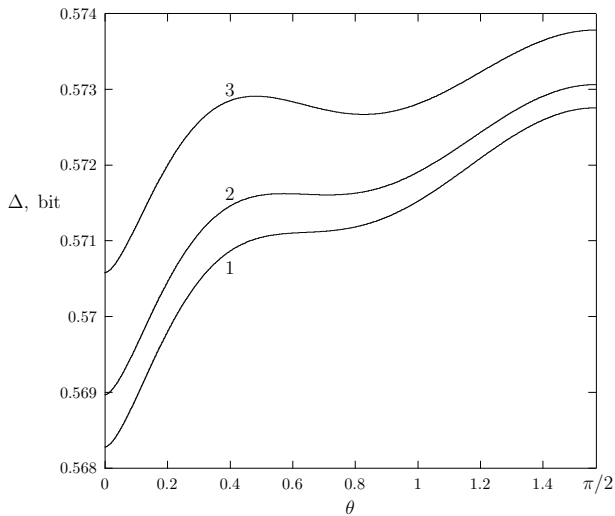,width=8cm}
\caption{
Measurement-dependent one-way quantum deficit $\Delta(\theta)$ along
the line $q_1+q_2=0.75$ by $q_1=0.72~(1)$, 0.72015~(2), and 0.7205~(3).
The bimodality appearing from an inflection point is clearly seen
}
\label{fig:zd0720-5}
\end{center}
\end{figure}
Consider now the behavior of post-measurement entropy and measurement-dependent one-way
deficit in the bulk area of ${\cal T}$.
We can inspect the total domain taking all possible straight-line trajectories
$q_1+q_2=const\le1$.
The behavior of the system is, obviously, symmetric relative to the middle of such
trajectories.
Take, for instance, the trajectory $q_1+q_2=0.75$ which is shown in Fig.~\ref{fig:zpd4}
by the straight line 3.
The shape of the curve $\Delta(\theta)$ has the monotonically increasing type in the middle
of this trajectory ($q_1=q_2=0.375$).
However, with the increase of the value of parameter $q_1$, the birth of a pair of extrema from
an inflection point occurs inside the interval $(0,\pi/2)$;
the situation is illustrated in Fig.~\ref{fig:zd0720-5}.
This phenomenon happens at the value of $q_1=0.72015$.
According to the definition (see, e.g., Ref.~\cite{V86}) a function having two
extrema in some interval is called bimodal on this interval.

\begin{figure}[t]
\begin{center}
\epsfig{file=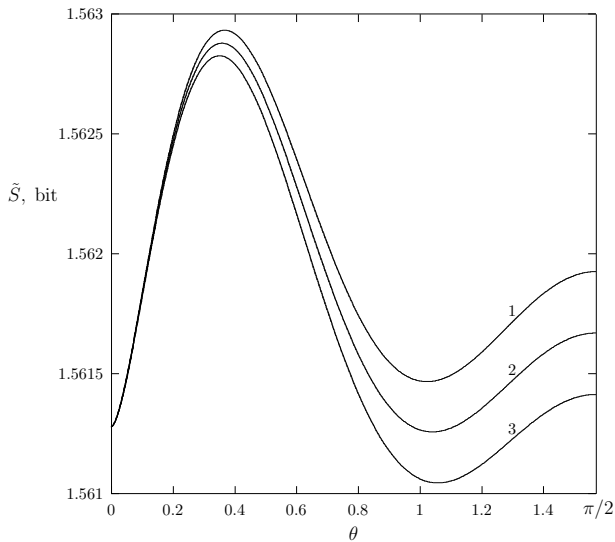,width=8cm}
\caption{
Post-measured entropy $\tilde S$ as a function of $\theta$
by $q_2=0.75-q_1$ and $q_1=0.7215~(1)$, 0.7216~(2), and 0.7217~(3).
}
\label{fig:zs1}
\end{center}
\end{figure}

With further increase of the $q_1$ value a qualitatively new effect is observed.
We demonstrate it by the curves $\tilde S(\theta)$ shown in Fig.~\ref{fig:zs1}.
When the parameter $q_1$ achieves the value of 0.72159,
the position of global minimum suddenly jumps through a finite step $\Delta\vartheta$
from zero to $\vartheta=1.0409\approx 60^{\circ}$ (see Fig.~\ref{fig:zs1}).
As a result, the fracture is arisen on the continuous curve
of minimized one-way quantum deficit ${\rm\Delta}(q_1)$. 
The position of the fracture point is determined from the equation
$\tilde S_0=\tilde S_\vartheta$ or
\begin{equation}
   \label{eq:frac}
   \Delta_0=\Delta_{\vartheta}.
\end{equation}
After this the interior minimum lies lower than another minimum located
at the endpoint $\theta=0$.
Notice that behavior of curve 3 in Fig.~\ref{fig:zs1} leads to a contradiction with
Eq.~(\ref{eq:D2}), i.e., the equation is incorrect for general X states.

\begin{figure}[t]
\begin{center}
\epsfig{file=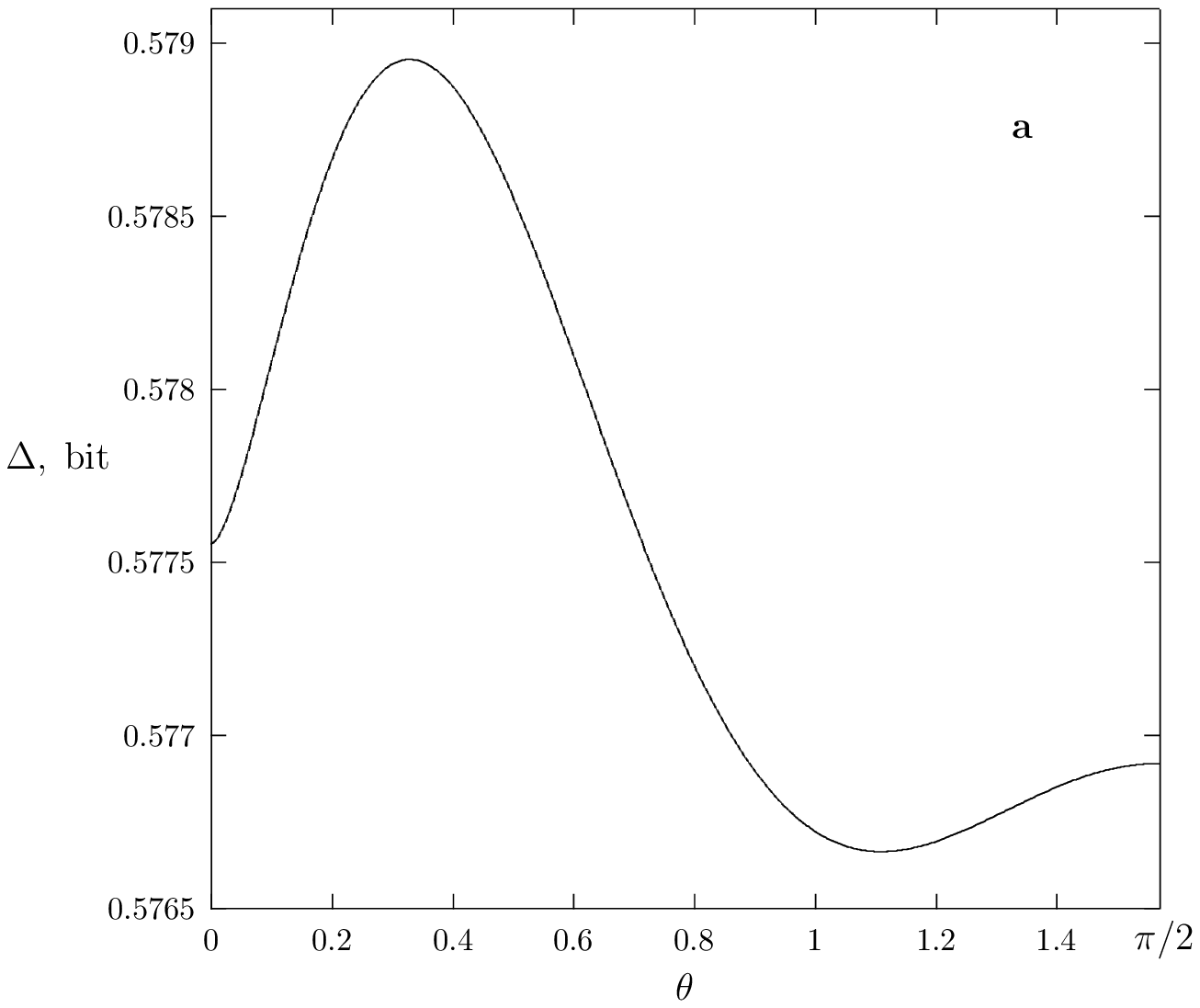,width=5.2cm}
\vspace{1cm}
\hspace{1cm}
\epsfig{file=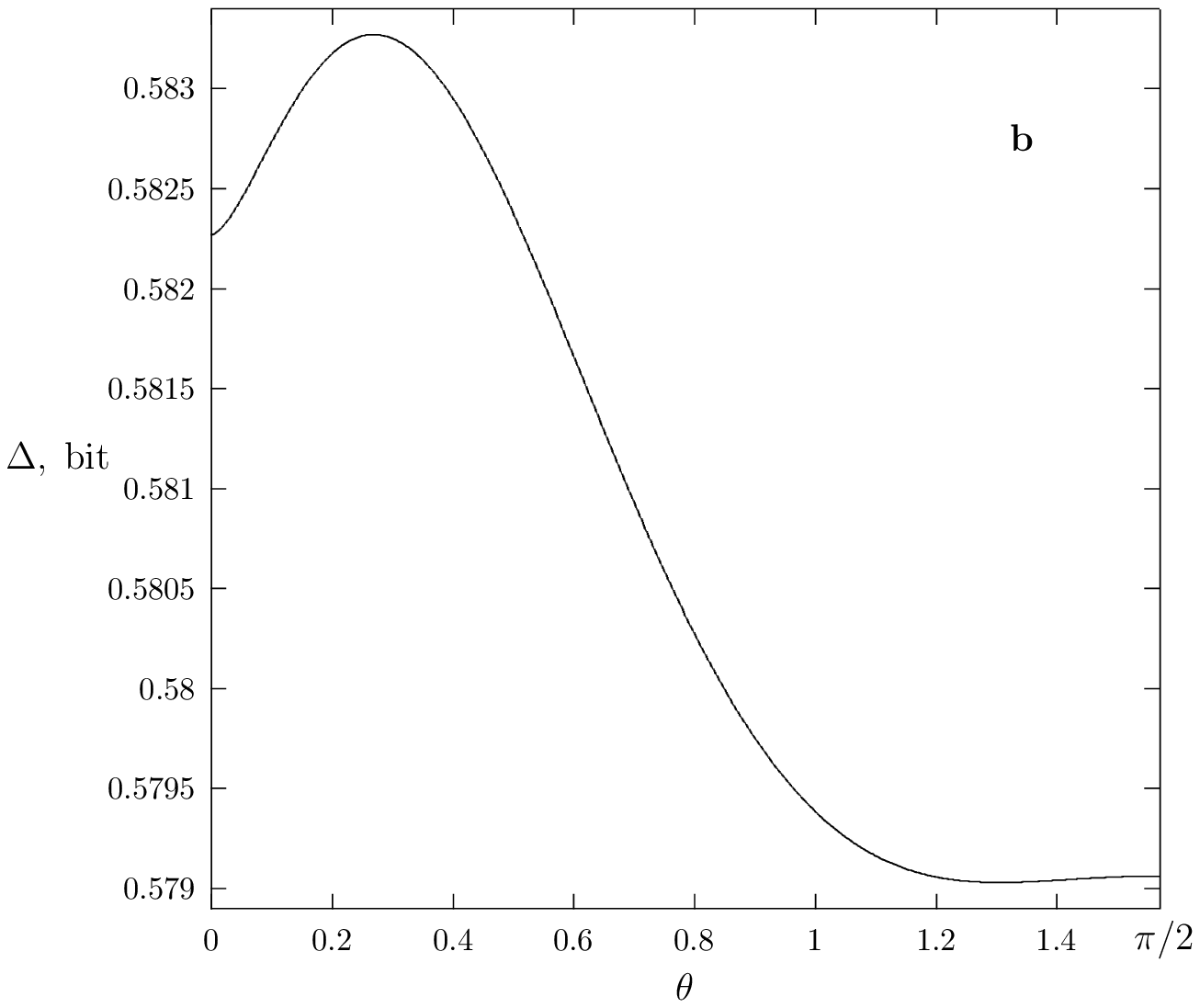,width=5.2cm}
\epsfig{file=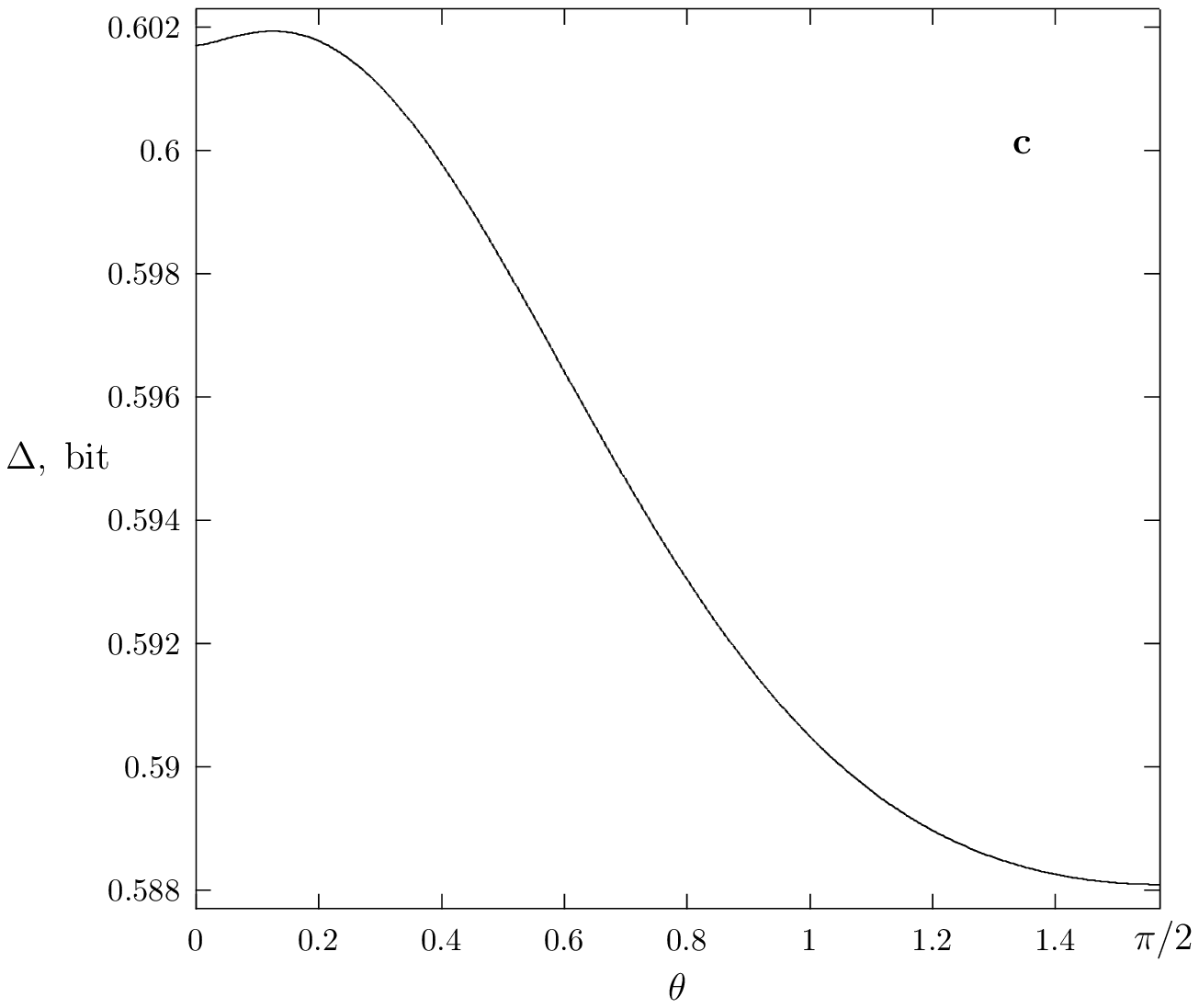,width=5.2cm}
\vspace{1cm}
\hspace{1cm}
\epsfig{file=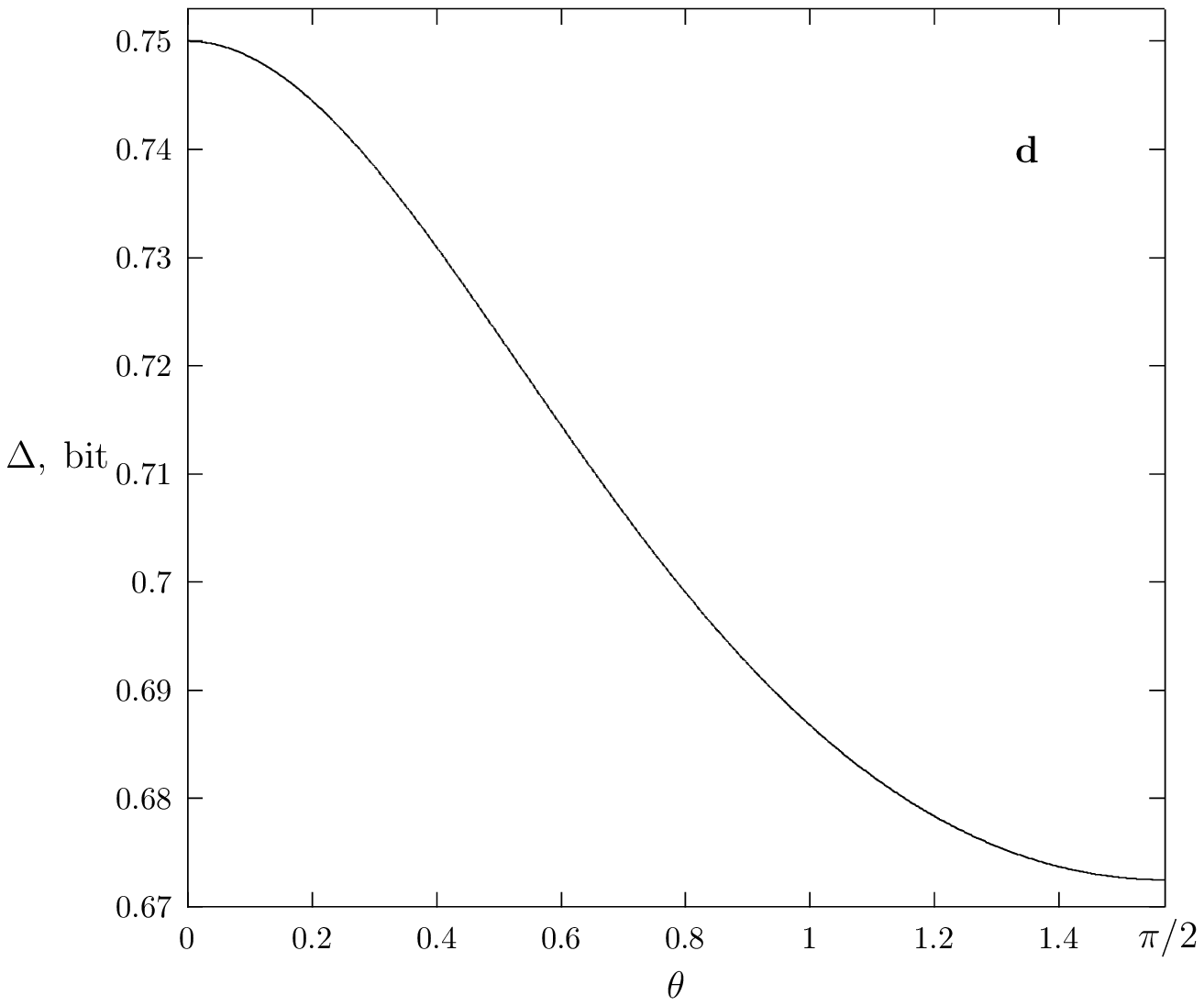,width=5.2cm}
\caption{
Measurement-dependent one-way quantum deficit $\Delta(\theta)$ along
the line $q_1+q_2=0.75$ by $q_1=0.722~(a)$, 0.723~(b), 0.727~(c), and 0.75~(d).
Minimum on the curve disappears at the endpoint $\theta=\pi/2$
through the bifurcation mechanism
whereas the maximum annihilates at the endpoint $\theta=0$
via the singularity mechanism
}
\label{fig:zd4}
\end{center}
\end{figure}
With further increasing $q_1$ the interior minimum smoothly moves to the point
$\theta=\pi/2$ and disappears at $q_1=0.72358$ when the trajectory crosses the curve 2,
i.e., the $\pi/2$-boundary (see Fig.~\ref{fig:zpd4}).
The dynamics of corresponding deformations of $\Delta(\theta)$ is depicted in
Fig.~\ref{fig:zd4}.
After crossing the $\pi/2$-boundary, the behavior of $\rm\Delta$ undergoes to the branch
$\Delta_{\pi/2}$ up to the point of contact of trajectory with the Cartesian axis,
i.e., up to $q_1=0.75$, where the interior maximum of $\Delta(\theta)$ disappears
at the endpoint $\theta=0$.
This happens through a new non-bifurcation (and non-inflection) mechanism.
Since the second derivative $\Delta^{\prime\prime}(\theta)$ at $\theta=0$ diverges
out of
the Cartesian axes we will call this mechanism the singular one.

\begin{figure}[t]
\begin{center}
\epsfig{file=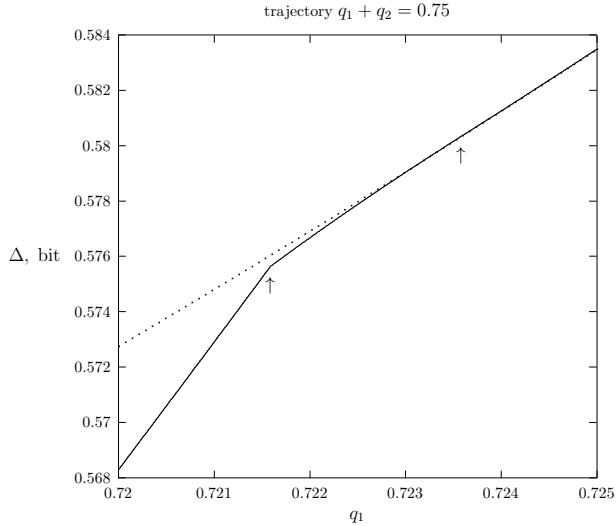,width=8cm}
\caption{
One-way quantum deficit $\rm\Delta$ vs $q_1$ along the path $q_1+q_2=0.75$
is shown by solid line.
Dotted line corresponds to the branch $\Delta_{\pi/2}$.
Fraction $\Delta_\vartheta$ with variable optimal measured angle lies between two arrows.
The transition $\Delta_0\leftrightarrow\Delta_\vartheta$ is displayed as a fracture on the curve
$\rm\Delta(q_1)$ whereas the $\Delta_\vartheta\leftrightarrow\Delta_{\pi/2}$ one is hidden ---
the curve is here continuous and smooth
}
\label{fig:zd075-1a}
\end{center}
\end{figure}

As a result, the one-way quantum deficit is obtained from the final equation
\begin{equation}
   \label{eq:DD}
   {\rm\Delta}=\min\{\Delta_0,\Delta_{\vartheta},\Delta_{\pi/2}\},
\end{equation}
where $\Delta_0$ and $\Delta_{\pi/2}$ are known in closed analytical forms and
$\Delta_{\vartheta}$ is found numerically.
The behavior of one-way deficit along the trajectory $q_1+q_2=0.75$ is shown
in Fig.~\ref{fig:zd075-1a}.

Either totally or partially similar behavior takes place for other trajectories
$q_1+q_2=const$ which go lower the intersection point of curves defined by equations
$\Delta_0=\Delta_{\pi/2}$ and $\Delta^{\prime\prime}(\pi/2)=0$, i.e, when
$const\le0.769\,095$.
For example, in the case of trajectory $q_1+q_2=0.65$, the bimodality appears
at $q_1\simeq0.631$ and a jump of optimal measurement angle from zero happens
at $q_1=0.631\,766$.
Values of jump angles $\Delta\vartheta$ in different cases are collected in
Table~\ref{tab:1}.
\begin{table}[t]
\caption{
Jumps of optimal measured angles, $\Delta\vartheta$, on the boundary
between the phases $\Delta_0$ and $\Delta_\vartheta$
\upshape\upshape}
\label{tab:1}
\begin{tabular}{lll}
\hline\noalign{\smallskip}
$q_1$ & $q_2$ & $\Delta\vartheta$ \\
\noalign{\smallskip}\hline\noalign{\smallskip}
$0.5$ & $0$ & $0=0^\circ$  \\
$0.544\,535$ & $0.55-q_1$ & $0.1267\approx7^\circ$ \\
$0.588\,104$ & $0.6-q_1$ & $0.2470\approx14^\circ$ \\
$0.631\,766$ & $0.65-q_1$ & $0.4020\approx23^\circ$ \\
$0.676\,082$ & $0.7-q_1$ & $0.6252\approx36^\circ$ \\
$0.721\,590$ & $0.75-q_1$ & $1.0409\approx60^\circ$ \\
$0.739\,409$ & $0.029\,686$ & $\pi/2=90^\circ$ \\
\noalign{\smallskip}\hline
\end{tabular}
\end{table}

\begin{figure}[t]
\begin{center}
\epsfig{file=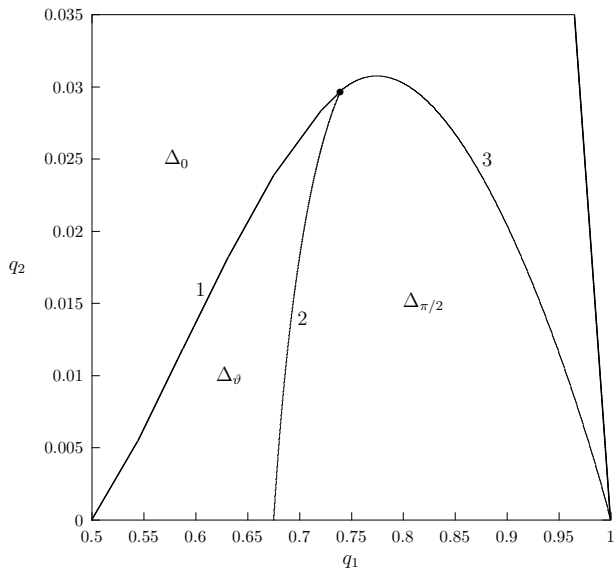,width=8cm}
\caption{
A fragment of phase diagram.
The boundary 1 is defined by equation $\Delta_0=\Delta_\vartheta$,
2 is the $\pi/2$-boundary, and the boundary 3 is defined by equation
$\Delta_0=\Delta_{\pi/2}$.
The black circle ($\bullet$) is the intersection point of $\pi/2$-boundary
with equilibrium curve of phases $\Delta_0$ and $\Delta_{\pi/2}$.
(This figure represents a part of the domain of definition shown in Fig.~\ref{fig:zpd4}.)
}
\label{fig:zpd6}
\end{center}
\end{figure}

A set of points where the optimal measurement angle discontinuously changes
from zero to a finite value gives the jumping (or hopping) boundary;
it serves instead of the absent ordinary 0-boundary (see Fig.~\ref{fig:zpd6}).
Between this boundary and the $\pi/2$-one, there exists an intermediate phase (fraction)
$\Delta_\vartheta$ with state-dependent optimal measurement angle $\vartheta$
which smoothly varies from some nonzero value to $\pi/2$.
The flat of two subregions with variable optimal angle, $\Delta_\vartheta$, is near $1\%$
of the one of total domain ${\cal T}$.

\begin{figure}[t]
\begin{center}
\epsfig{file=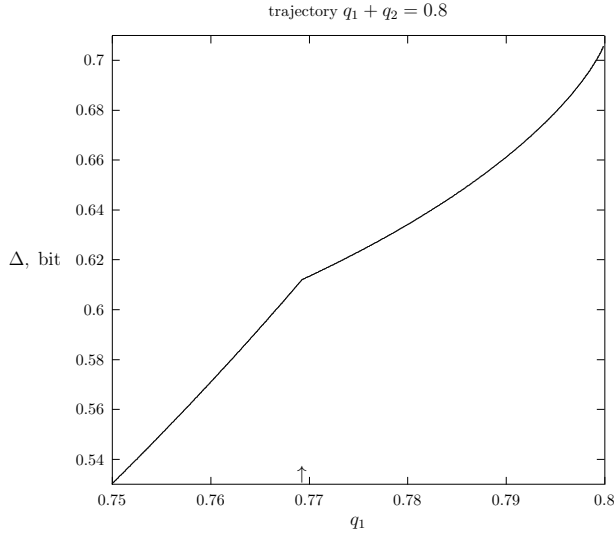,width=8cm}
\caption{
Dependence of $\rm\Delta$ vs $q_1$ by $q_2=0.8-q_1$.
Arrow marks the position of a fracture at the point $q_1=0.769\,269$, where the one-way deficit
undergoes from the branch $\Delta_0$ to the $\Delta_{\pi/2}$ one
}
\label{fig:zd08a}
\end{center}
\end{figure}

When $const>0.769\,095$ (i.e., when the trajectories lie above the black circle shown
in Fig.~\ref{fig:zpd6}), the situation is different.
With increasing $q_1$ from middle values to the endpoint on the axis $Oq_1$ the curves
$\tilde S(\theta)$ or $\Delta(\theta)$ are deformed from monotonically increasing shape to the shape
with a single interior maximum (which is born at the point, where
$\Delta^{\prime\prime}(\pi/2)=0$) and then a sudden transition $\Delta_0\to\Delta_{\pi/2}$
occurs at the boundary defined by the relation $\Delta_0=\Delta_{\pi/2}$
(line~3 in Fig.~\ref{fig:zpd6}).
Here, there is no intermediate region $\Delta_\vartheta$ and
the transition $\Delta_0\to\Delta_{\pi/2}$ is characterized visually by a fracture
on the curve ${\rm\Delta}(q_1)$.
Typical behavior of one-way deficit is shown in Fig.~\ref{fig:zd08a} along the trajectory
$q_1+q_2=0.8$.

So, the presented method to calculate the one-way quantum deficit of X states
is reduced first of all to careful analyzing of the shapes of post-measured entropy
or measurement-dependent one-way deficit curves.
One should also solve equations for the boundaries between three possible phases (branches):
Eqs.~(\ref{eq:D0D1}), (\ref{eq:D11}), and (\ref{eq:frac}).
After this the one-way quantum deficit is obtained from the piecewise-analytical-numerical
formula (\ref{eq:DD}).

\section{Summary and concluding remarks}
\label{sect:Concl}
In this paper we have found that besides the monotonic and unimodal behavior
the post-measured entropy and hence the measurement-dependent one-way quantum deficit
upon the measurement angle can have a new kind of behavior.
Namely, these functions can exhibit the bimodal shape in the open interval $(0,\pi/2)$
for different regions in the space of X state parameters.
This  expands the variety of behavior for the one-way quantum deficit $\rm\Delta$.
In particular, a new state-dependent phase (fraction) which is characterized by
a {\em partial} interval of optimal measured angles has been found.
Instead of smooth conjugation of the branches $\Delta_0$ and $\Delta_{\pi/2}$ this leads to
a fracture on the curve of one-way deficit.

New mechanism of a boundary arising between the phases via jumping
the optimal measured angle on a finite step has been discovered.
Instead of bifurcation conditions (\ref{eq:D11}) the boundary is now determined by
a relation like (\ref{eq:frac}).
The study of post-measured entropy shapes is the general way to determine the correct
one-way quantum deficit.

This is in contrast with the behavior of conditional entropy and, consequently,
measurement-dependent quantum discord in the same regions of parameter space:
their behavior is restricted by monotonic and unimodal types.
In any case, this rather simple and therefore attractive picture is valid for the
different specific cases and subclasses of X states \cite{Y14,Y15,Y17}.
In particular, such a behavior of conditional entropy is confirmed for the
symmetric XXZ states \cite{Y17} those may be written in an equivalent form as
\begin{equation}
   \label{eq:rho3}
   \rho_{AB}=q_1|\Psi^+\rangle\langle\Psi^+| + q_2|\Psi^-\rangle\langle\Psi^-|
	 + q_3|00\rangle\langle 00|
	 + q_4|11\rangle\langle 11|
\end{equation}
with $q_1+q_2+q_3+q_4=1$.

An intriguing question remains: are there any more general shapes of curves
for the post-measured entropy of X states?
For instance, can this entropy have trimodal and, maybe, multimodal dependence?
The answer to these and other questions should come from the future investigations
of post-measurement entropy shapes in the full five-parameter X-state space.


\vspace{-10mm}
\section*{}
{\bf Acknowledgment}\ 
The work was supported by the Russian Foundation for Basic Research.




\end{document}